\documentclass[aps,prb,twocolumn,longbibliography,superscriptaddress]{revtex4-1}

\usepackage[T1]{fontenc}
\usepackage{soul}
\usepackage{amssymb}
\usepackage{amsmath}

\usepackage{latexsym}
\usepackage{graphicx} 
\usepackage{epstopdf}
\usepackage{comment}
\usepackage{bbold}

\usepackage{color}
\definecolor{aurometalsaurus}{rgb}{0.43, 0.5, 0.5}
\definecolor{mfcolor}{rgb}{0.0, 0.4, 1.0}
\definecolor{mzcolor}{rgb}{0.2, 0.7, 0.0}

\begin{document}

\title{Unconventional topological superconductivity and phase diagram for an effective two-orbital model as applied to twisted bilayer graphene}

\author{M. Fidrysiak}
 \email{maciej.fidrysiak@uj.edu.pl}
\affiliation{Marian Smoluchowski Institute of Physics, Jagiellonian University, ul. {\L}ojasiewicza 11, 30-348 Krak{\'o}w, Poland }
 \author{M. Zegrodnik}
 \email{michal.zegrodnik@agh.edu.pl}
\affiliation{Academic Centre for Materials and Nanotechnology, AGH University of Science and Technology, \\ Al.~Mickiewicza~30,~30-059~Krak\'{o}w,~Poland}
 \author{J. Spa{\l}ek}%
 \email{jozef.spalek@uj.edu.pl}
\affiliation{Marian Smoluchowski Institute of Physics, Jagiellonian University, ul. {\L}ojasiewicza 11, 30-348 Krak{\'o}w, Poland }


\begin{abstract}
We consider the superconducting and Mott-insulating states for the twisted bilayer graphene, modeled as two narrow-band system of electrons with appreciable intraatomic Coulomb interactions. The interaction induces kinetic exchange which leads to real-space, either triplet- or singlet-spin pairing, in direct analogy to heavy-fermions and high-temperature superconductors. By employing the statistically-consistent Gutzwiller method, we construct explicitly the phase diagram as a function of electron concentration for the spin-triplet $d_{x^2 - y^2}+id_{xy}$ paired case, as well as determine the topological edge states. The model reproduces principal features observed experimentally in a semi-quantitative manner. The essential role of electronic correlations in driving both the Mott-insulating and superconducting transitions is emphasized. The transformation of the spin-triplet state into its spin-singlet analogue is also analyzed, as well as the appearance of the phase separated superconducting+Mott-insulating state close to the half filling.
\end{abstract}

\maketitle

\section{Introduction}
\label{sec:introduction}
The nature of electronic states and, in particular, the microscopic mechanism of   \textit{unconventional} pairing in strongly correlated matter is one of the fundamental problems in condensed matter physics. This is because in the systems such as heavy fermions, \cite{PfleidererRevModPhys2009} high-temperature superconductors (SC), \cite{HottBook2015} or selected atomic systems in optical lattice,\cite{LewensteinBook2012} the interparticle interaction energy can exceed by far the single-particle (kinetic, band) energy. In that situation, specific phenomena induced by the interelectronic correlations occur, such as the Mott or the Mott-Hubbard localization,\cite{GebrahrdBook1997} unconventional superconductivity (SC) associated with real-space pairing,\cite{OgataRepProgPhys2008} as well as specific magnetic behavior, such as metamagnetic transition to localized state, spin-dependent masses of quasiparticles,\cite{SpalekPhysStatSolidiB2006} and quantum critical behavior.\cite{CarrBook2011}

In this context, the recent discovery of SC and the concomitant Mott insulating (MI) behavior \cite{CaoNature2018_SC,CaoNature2018_MI,ChenArXiV2018} of twisted bilayer graphene (TBG) provides a model situation for studying phenomena ascribed to high-temperature SC.  This is because TBG represents a truly two-dimensional system and the ratio of interaction amplitude to the Fermi energy can be varied experimentally in a controlled manner by applying gate voltage to the sample substrate. What is particularly important here is that the behavior close to the SC-MI boundary can be studied systematically by changing the carrier concentration, without introducing the ubiquitous atomic disorder, as is the case in the high-temperature SC. Also, the whole concentration-dependent phase diagram can be sampled by changing the gate voltage and hence, the data represent intrinsic system properties. However, there are also differences with respect to the high-$T_c$ SC. The main difference is that TBG is an inherently multi-orbital system,\cite{YuanArXiv2018} whereas the high-$T_c$ SC may be mapped onto single-orbital models.\cite{ZegrodnikPhysRevB2017} This circumstance constitutes a basis for an extension of our single-band version of SGA  method statically-consistent Gutzwiller method (SGA)\cite{JedrakPhysRevB2011,ZegrodnikNewJPhys,Abram2017} to the present situation.

Here we start from an effective two-band Hubbard model with substantial intra- and inter-band interactions,  placed on triangular-lattice of moir{\'e} type and solve it explicitly within SGA. Initially, we assume full $\mathrm{SU(4)}$ symmetry in the spin-orbital space and subsequently extend the model by incorporating the effects of the symmetry-breaking Hund's rule coupling. The resulting second-order kinetic exchange interaction may lead to either spin-triplet or spin-singlet pairing with the increasing electron concentration. On this basis, we compose the phase diagram encompassing the SC and MI phases, as well as a phase-separation regime in between SC and MI states. Technical details of the discussion, as well as the extended analysis of the diagrammatic extension of the SGA treatment, are deferred to Appendices~\ref{appendix:exchange_integrals}-\ref{appendix:diagrams}.

\section{Model and real-space pairing: \textrm{SU(4)} scenario}
\label{sec:model}

We consider a two-orbital model ($l=1, 2$) on a triangular lattice for which a single site represents one moir\'e unit cell. The two orbitals correspond to the two original valleys at the Brillouin zone corners. The starting $\mathrm{SU(4)}$ symmetric Hamiltonian is\cite{BalentsArXiv2018}

\begin{align}
  \label{eq:hamiltonian}
  \mathcal{H} = & t \sum\limits_{\langle i, j \rangle} \left(c^{(l)\dagger}_{i\sigma} c^{(l)}_{j\sigma} + \mathrm{H.c.}\right) + \frac{U}{2} \sum\limits_{i l} \left(\hat{n}_{i}^{(l)}\right)^2 \nonumber \\ &+ \frac{U^\prime}{2} \sum\limits_{i l} \hat{n}_i^{(l)} \hat{n}_i^{(\bar{l})} - \frac{U}{2} \sum\limits_{il} \hat{n}_{i}^{(l)},
\end{align}

\noindent
where $t$, $U$, and $U^\prime$ denote the hopping, and intraorbital and interorbital Coulomb interactions, respectively. The operator $c^{(l)\dagger}_{i\sigma}$ creates an electron with spin $\sigma$ on orbital $l$ at site $i$ and $\hat{n}_i^{(l)} \equiv \sum_\sigma c_{i\sigma}^{(l)\dagger} c_{i\sigma}^{(l)}$ is the orbital particle-number operator (we use the notation $\bar{l} = 2, 1$ for $l = 1, 2$, respectively). Hereafter we assume approximate $\mathrm{SU}(4)$ spin-orbital symmetry by taking $U = U'$. This means that the Hund's rule coupling is disregarded at this point (see below). Also, the interlayer hopping $t^{12}$ is neglected, since it can be shown that with such an orbitally independent form of hybridization $\sim t^{12}$ term can be incorporated into an effective canonical structure without it.\cite{SpalekSolStateCommun1985}

The dominant paring channels  can be identified by referring to canonical perturbation expansion\cite{SpalekSolStateCommun1985,SpalekJPhysC1980} in the manner analogous to that for the one-band Hubbard model. In the simplest case of $U = U^\prime$, the interaction takes the form $\mathcal{H}_\mathrm{int} = U/2 \times \sum_i \hat{n}_i (\hat{n}_i - 1)$, where $\hat{n}_i \equiv \hat{n}_i^{(1)} + \hat{n}_i^{(2)}$ is the total particle number operator for lattice site $i$. The configurations with large local density of electrons are thus disfavored by the interaction, and hopping processes generating such states should be eliminated by means of the canonical transformation. This procedure (cf. Appendix~\ref{appendix:exchange_integrals}), after employing standard approximations, leads to the kinetic exchange taking a general functional form

\begin{align}
  \label{eq:exchange}
  \mathcal{H}_\mathrm{ex} \sim & -J_\mathrm{ex} {\sum \limits_{i j }}' \sum \limits_{\sigma \sigma^\prime l l^\prime} c^{(l)\dagger}_{i\sigma} c^{(l)}_{j\sigma} c^{(l')\dagger}_{j\sigma'} c^{(l')}_{i\sigma'} \nonumber \\ & = -J_\mathrm{ex} {\sum \limits_{i j }}' \sum \limits_{\sigma \sigma^\prime l l^\prime} c^{(l)\dagger}_{i\sigma}  c^{(l')\dagger}_{j\sigma'} c^{(l')}_{i\sigma'} c^{(l)}_{j\sigma} - J_\mathrm{ex} {\sum \limits_{i j }}' \hat{N}_i \nonumber \\ & = \frac{J_\mathrm{ex}}{4} {\sum \limits_{i j }}' \sum \limits_{\alpha\beta}  c^{\dagger}_{j} (\sigma^\alpha)^T (\tau^\beta)^T  (c^{\dagger}_{i})^T \times (c_{i})^T \sigma^\alpha \tau^\beta c_{j} - \nonumber \\ & - J_\mathrm{ex} {\sum \limits_{i j }}' \hat{N}_i,
\end{align}

\noindent
where $\sigma^\alpha$ and $\tau^\alpha$ are Pauli matrices acting on the spin- and orbital- indices, respectively (the summation is performed over $\alpha = 0, \ldots, 3$, with $\sigma^0 \ = \tau^0 \equiv \mathbb{1}$). We have used the compact notation $c^\dagger \equiv (c^{(1)\dagger}_\uparrow, c^{(1)\dagger}_\downarrow, c^{(2)\dagger}_\uparrow, c^{(2)\dagger}_\downarrow)$, $T$ denotes transposition, and $J_\mathrm{ex}$ sets the effective kinetic exchange scale $O(t^2/U)$. The primed symbols ${\sum}'$ means that  summation is performed over nearest neighbors. Note now that, for $(\sigma^\alpha)^T (\tau^\beta)^T = - \sigma^\alpha \tau^\beta$, an additional minus sign is generated from the $c^{\dagger}_{j} (\sigma^\alpha)^T (\tau^\beta)^T  (c^{\dagger}_{i})^T$ term by performing the transposition, rendering the interaction attractive in some pairing channels. This occurs for spin-singlet, orbital-triplet ($\alpha = 2$, and $\beta = 0, 1, 3$) and for spin-triplet, orbital-singlet ($\alpha = 0, 1, 3$, and $\beta = 2$) cases. All other pairing symmetries are disfavored. Hereafter we adopt the point of view that an additional Hund's rule interorbital interaction that breaks full $\mathrm{SU}(4)$ symmetry, which is not explicitly included in the original Hamiltonian~\eqref{eq:hamiltonian}, would tip the balance towards the spin-triplet, orbital-singlet counterpart, in broad doping range $0 \leq n \leq 2$ (cf. Appendix~\ref{appendix:exchange_integrals}). Analogous conclusion was previously drawn in Ref.~\citenum{BalentsArXiv2018}. The approximate $\mathrm{SU}(4)$ symmetry implies, however, that those states should be energetically close to each other.

An important remark is in place here. In general, the spin-singlet pairing should appear as the system approaches half-filling, since then the intraorbital antiferromagnetic kinetic exchange becomes dominant (cf. Appendix~\ref{appendix:exchange_integrals} and Refs.~\citenum{ChaoPhysStatSolidiB1977,SpalekJPhysC1980}). However, it can be shown that upon changing the sign on $J$, the formalism for the spin-triplet case formally coincides with that for the spin-singlet situation. Specifically, whereas here we explicitly consider here solely the spin-triplet case, the reported results are also equally valid to the spin-singlet scenario by an appropriate unitary transformation (cf. Appendix~\ref{appendix:singlet-triplet_transformation} and Ref.~\citenum{ZegrodnikPhysRevB2014}). In either case, the parity of the SC order parameter is even. Previous studies of the triangular-lattice Hubbard model \cite{ChenPhysRevB2013} suggest that, in this geometry, unconventional $d + id$ symmetry might be realized in order to optimize the condensation energy. This is due to the fact that, contrary to the usual $d$-wave pairing, entire Fermi surface becomes then gapped upon the SC transition. We thus consider first the case of the $A$-type, i.e., spin-triplet, orbital-singlet, $d + id$ pairing, defined by the following relations between SC amplitudes

\begin{align}
  &\langle c_{i\sigma}^{(1)} c_{j\sigma}^{(2)} \rangle = -\langle c_{i\sigma}^{(2)} c_{j\sigma}^{(1)} \rangle \text{\hspace{1em}(orbital-singlet)},    \label{eq:sc_structure_singlet} \\ & \langle c_{i\sigma}^{(1)} c_{j\sigma}^{(2)} \rangle = \langle c_{i\bar{\sigma}}^{(1)} c_{j\bar{\sigma}}^{(2)} \rangle \text{\hspace{1em}($A$-phase)},  \label{eq:sc_structure_a_phase} \\ & \langle c_{i\sigma}^{(1)} c_{(R_i(\theta)j)\sigma}^{(2)} \rangle = \exp\left(2 i \theta\right) \langle c_{i\sigma}^{(1)} c_{j\sigma}^{(2)} \rangle \text{\hspace{1em}($d+id$)},  \label{eq:sc_structure_d}
\end{align}

\noindent
where $R_i(\theta)$ is the rotation by the angle $\theta$ around the axis perpendicular to the lattice plane, and going through the site $i$. The even-parity property follows from the condition~\eqref{eq:sc_structure_d} for $\theta = \pi$ and translational symmetry. It should be noted that an alternative approach of purely real extended $s$-wave pairing in TBG has been presented very recently, \cite{RayarXiV2018} where the Eliashberg formalism has been used within which the pairing is induced due to the many-body spin- and charge-fluctuations. That paper represents a complementary weak-correlation perspective.

\begin{figure}
  \includegraphics[width = 1.0\columnwidth]{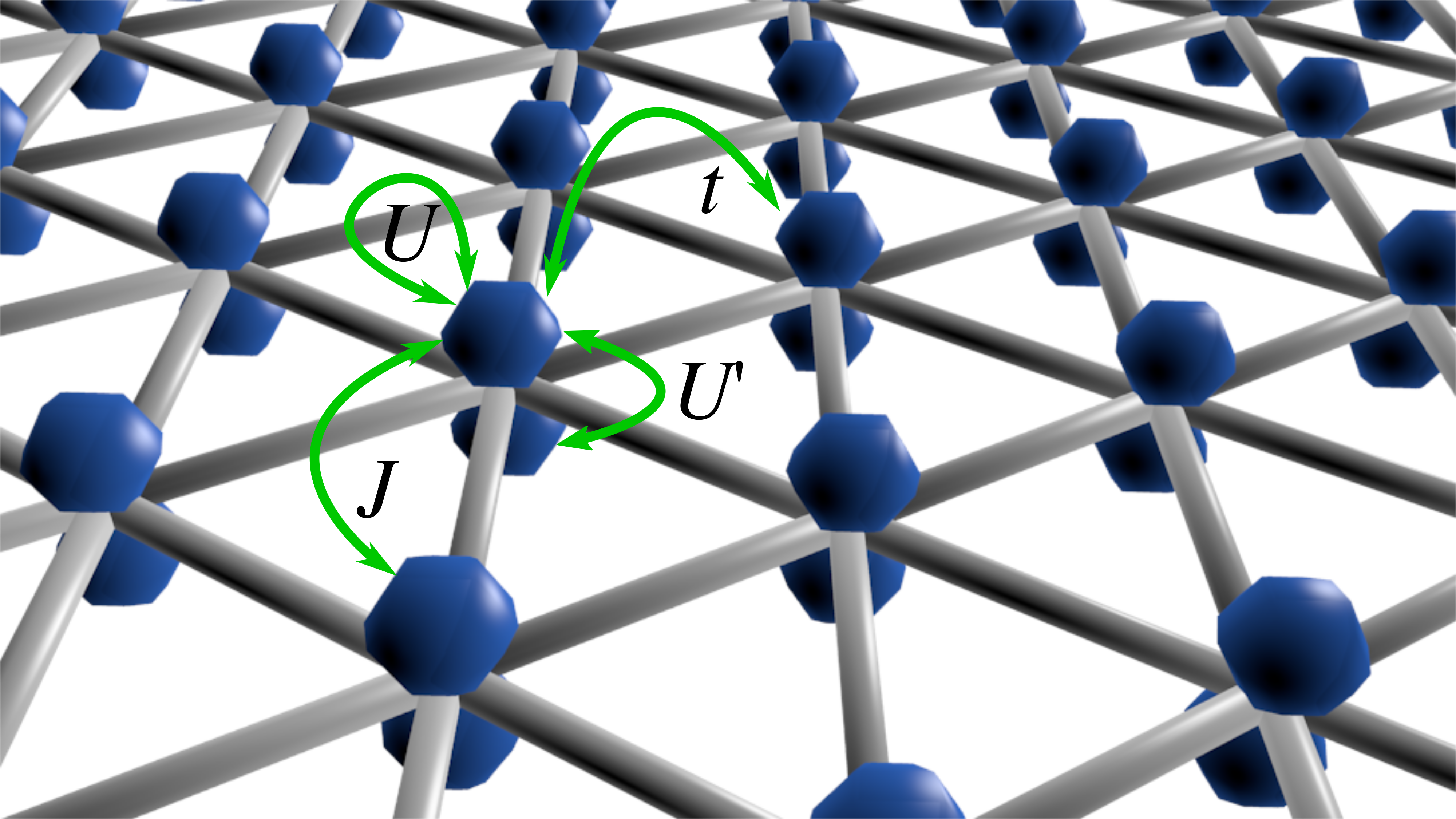} %
  \caption{Illustration of the effective two-band model of twisted bilayer graphene. The green arrows mark intraorbital repulsion $U$, interorbital repulsion $U'$, intersite exchange interaction $J$, as well as the single-particle hopping integral $t$.}
  \label{fig:lattice}
\end{figure}

To investigate this scenario, instead of employing the original model \eqref{eq:hamiltonian} with the general kinetic exchange \eqref{eq:exchange}, we resort to a simpler effective Hamiltonian favoring spin-triplet pairing, defined as

\begin{align}
  \label{eq:hamiltonian_final}
  \mathcal{H}_\mathrm{eff} = & t \sum\limits_{\langle i, j \rangle} \left(c^{(l)\dagger}_{i\sigma} c^{(l)}_{j\sigma} + \mathrm{H.c.}\right) + \frac{U}{2} \sum\limits_{i l} \left(\hat{n}_{i}^{(l)}\right)^2 \nonumber \\ &+ \frac{U^\prime}{2} \sum\limits_{i l} \hat{n}_i^{l} \hat{n}_i^{(\bar{l})} - \frac{U}{2} \sum\limits_{il} \hat{n}_{i}^{(l)} - J \sum \limits_{\langle i, j\rangle} \hat{\mathbf{S}}_i \hat{\mathbf{S}}_j,
\end{align}

\noindent
where $\hat{\mathbf{S}}_i$ is the total-spin operator on lattice site $i$. The effective pairing coupling $J$ has been introduced.  The Hamiltonian \eqref{eq:hamiltonian_final} reproduces correctly attractive interaction in the spin-triplet channel and thus is applicable as long as solely paramagnetic and superconducting state of symmetry defined by Eqs.~\eqref{eq:sc_structure_singlet}-\eqref{eq:sc_structure_d} are considered. The effective model is illustrated in Fig.~\ref{fig:lattice}. A more detailed analysis of kinetic-exchange-integral in the two-band situation in discussed in Appendix~\ref{appendix:exchange_integrals}.

\section{Solution and phase diagram}
\label{sec:solution}

We employ now the SGA which has proven to be effective for various classes of correlated electron systems, including the high-$T_c$ cuprates \cite{JedrakPhysRevB2011} and spin-triplet ferromagnetic SC. \cite{ZegrodnikNewJPhys,KadzielawaArXiv2017} At zero temperature, SGA reduces to optimization of the ground state energy $E_G \equiv \langle \Psi_G | \mathcal{H} | \Psi_G \rangle / \langle \Psi_G | \Psi_G \rangle$ within the class of trial wave functions $|\Psi_G\rangle \equiv P_G |\Psi_0\rangle$. Here $|\Psi_0\rangle$ denotes a Slater determinant (describing uncorrelated electrons) and $P_G \equiv \prod_i P_{Gi}$ is a product of local correlators $P_{Gi} \equiv \lambda_0 |0\rangle_i{}_i\langle0| + \lambda_\uparrow |\uparrow \rangle_i{}_i\langle \uparrow| + \lambda_\downarrow |\downarrow \rangle_i{}_i\langle \downarrow| + \lambda_{\uparrow\downarrow} |\uparrow\downarrow \rangle_i{}_i\langle \uparrow\downarrow|$ that modify the local many particle electronic configurations by means of variational coefficients $\lambda_\gamma$. This allows to include the effect of strong correlations on top of the renormalized quasiparticle picture. Additional details concerning specific features of SGA and the estimates of higher order contributions obtained within the related diagrammatic approach \cite{ZegrodnikPhysRevB2017,ZegrodnikNewJPhys,Abram2017} are provided in Appendix~\ref{appendix:diagrams}.

Since the width of the narrow bands arising in the magic-angle graphene is $W \sim 10\,\mathrm{meV}$,\cite{CaoNature2018_SC,CaoNature2018_MI} the role of the local correlations is expected to be crucial. To consider this scenario in detail, we fix the parameters as $t = -3\,\text{-}\,5\,\mathrm{meV}$, $U = U' = 18|t|$, and $J = -|t|$. The calculations are performed at small nonzero temperature $T = 10^{-4}|t|/k_B$ for numerical purposes. This value maps onto absolute temperature scale of less than $6\,\mathrm{mK}$.

\begin{figure}[h!]
  \includegraphics[width = \columnwidth]{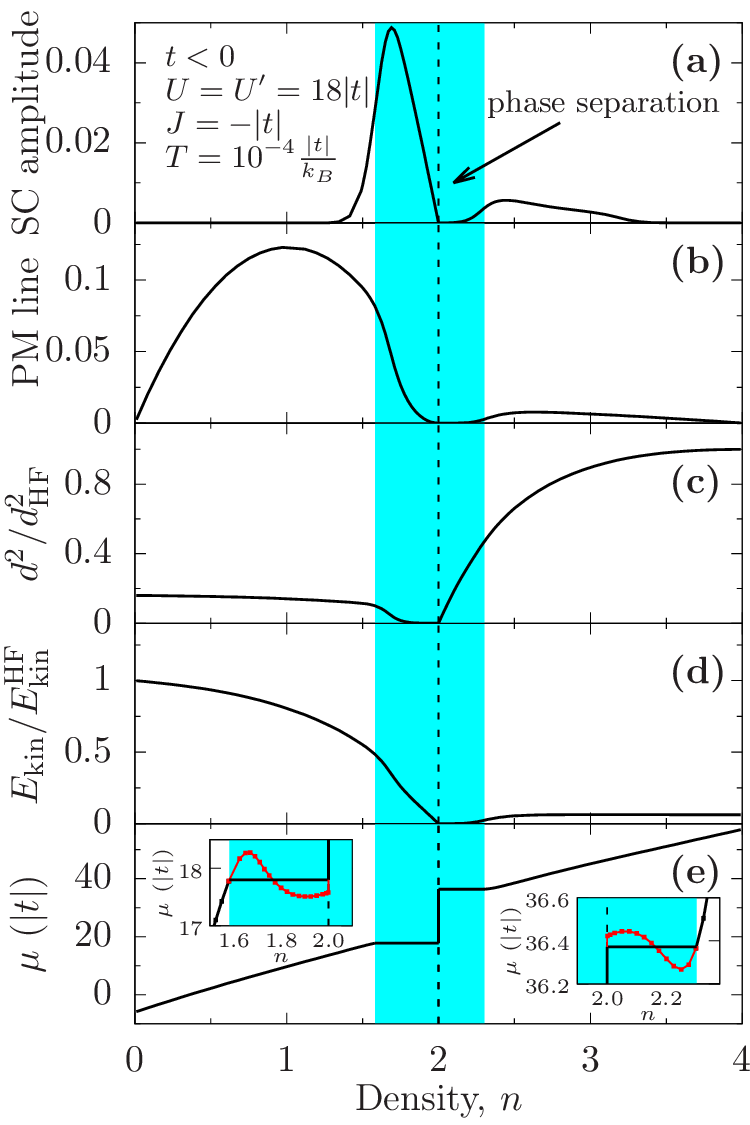} %
  \caption{Phase diagram for $t  < 0$, $U = U' = 18|t|$, $J=-|t|$, and temperature $T = 10^{-4}|t|/k_B$, obtained for $512 \times 512$ lattice. (a) Doping dependence of the dimensionless superconducting gap amplitude component $\langle c^{(1)}_{i\uparrow} c^{(2)}_{j\uparrow} \rangle_G$. The shaded area marks the phase-separation region, where the superconducting state appears separated spatially from the Mott insulating phase emerging near the half-filling. (b) Hopping probability $\langle c^{(l)\dagger}_{i\sigma} c^{(l)}_{j\sigma} \rangle_G$ which represents the electron itineracy. (c) Probability $d^2$ of the double site occupancy, normalized to its Hartree-Fock value $d^2_\mathrm{HF}$. (d) The ratio of kinetic energies calculated within the SGA and Hartree-Fock approximations. (e) Chemical potential $\mu$ as a function of electron density. Note that $\mu$ is constant in the phase-separation region and its value is determined by Maxwell construction. The close-ups of the phase-separation regions below and above the half-filling are displayed in the insets. The squares are computational data points and the red solid lines mark unstable spatially homogeneous solutions.}
  \label{fig:phase_diagram}
\end{figure}%

The calculated phase diagram for the model~\eqref{eq:hamiltonian_final} as a function of electron concentration $n$ per superlattice site is shown in Fig.~\ref{fig:phase_diagram}. In panel (a) we display the correlated SC amplitude component $\langle c^{(1)}_{i\sigma} c^{(2)}_{j\sigma}\rangle_G$ which is one of the principal results of the present contribution. Around the half-filling ($n = 2$), we obtain two asymmetric SC domes, with stronger SC correlations on the lower-concentration side of the phase diagram. Such an asymmetry is expected as triangular lattice is not bipartite and the electron-hole symmetry is explicitly broken. We point out a small, barely visible hump in SC amplitude, emerging near the integer filling $n = 3$, where the correlations are again enhanced. This feature is fairly weak for the present choice of parameters and is obscured by the larger dome closer to $n = 2$. The obtained two-dome structure exhibits a remarkable agreement with the recent experimental data for the magic-angle bilayer graphene.\cite{CaoNature2018_SC} We emphasize that the agreement is semi-quantitative as when the dimensionless SC amplitude is scaled by the characteristic energy $|t| \sim 3\,\text{-}\,5\,\mathrm{meV}$, the maximal gap parameter $\Delta/k_B \sim |t| \langle c^{(1)}_{i\sigma} c^{(2)}_{j\sigma}\rangle_G /k_B = 0.7\text{-}1.2\,\mathrm{K}$ matches well the measured critical temperatures. The shaded area in the phase diagram marks the phase-separation region, where the SC state coexists with an increasing fraction of the Mott insulating phase as the half-filling is approached. We point out that, even though the maximum of SC correlations for $n < 2$ lies within the phase separation regime, the sample-averaged SC amplitude still exhibits a two-dome structure with a proper particle-hole asymmetry. Panel (b) details doping evolution of the hopping probability, measured by the nearest-neighbor correlation function $\langle c^{(l)\dagger}_{i\sigma} c^{(l)}_{j\sigma} \rangle_G \approx q^2 \langle\Psi_0| c^{(l)\dagger}_{i\sigma} c^{(l)}_{j\sigma} |\Psi_0\rangle$, where $q$ is the renormalization factor. For kinematic reasons, these correlations drop to zero for an empty and completely filled system ($n = 0$ and $n = 4$, respectively). Due to strong correlations, this happens also close to the half-filling, where the Mott transition is approached. As is shown in panel (c), the probability of double occupancies $d^2 \equiv \langle \hat{n}_{i\uparrow}^{(l)} \hat{n}_{i\downarrow}^{(l)} \rangle_G$ normalized to its uncorrelated (Hartree-Fock) value $d^2_\mathrm{HF} = \langle{n}_{i\uparrow}^{(l)}\rangle_G \times  \langle{n}_{i\downarrow}^{(l)}\rangle_G$ is reduced accordingly. To complete the discussion of the electronic correlations, in Fig.~\ref{fig:phase_diagram}(d) we plot the ratio of renormalized to uncorrelated kinetic energy which roughly describes the mass enhancement $m/m^{*}$. As expected, the kinetic energy is suppressed as the Mott transition is approached. Note that the triangular lattice model, considered here, is magnetically frustrated and thus the long-range antiferromgnetic state may not be favored close to the Mott state. This justifies disregarding the magnetic ordering at the present stage of analysis.

Next, we discuss the phase separation occurring in our model. Figure~\ref{fig:phase_diagram}(e) shows the doping-dependence of the chemical potential. As the half-filling is approached from either above or below, the chemical potential eventually levels off as a function of electron concentration, to become extremely steep in close vicinity of $n = 2$. This behavior is reminiscent of that observed for the one-band Hubbard model,\cite{TandonPhysRevLett1999,KotliarPhysRevLett2002,EcksteinPhysRevB2007,SpalekPhysStatSolB1981,HubbardProcRoySocLondon1964} where phase separation occurs as well. The value of the chemical potential in this regime is determined by Maxwell construction, which is illustrated in the insets. The squares are computational data points and red color marks unstable solutions. At present, we are unable to numerically approach part of the curve sufficiently close to $n = 2$ to observe the full S-shaped function $\mu(n)$. The jump in $\mu$ for $n=2$ is a signature of the opening of the Hubbard gap.\cite{HubbardProcRoySocLondon1964}

\begin{figure}
  \includegraphics[width = \columnwidth]{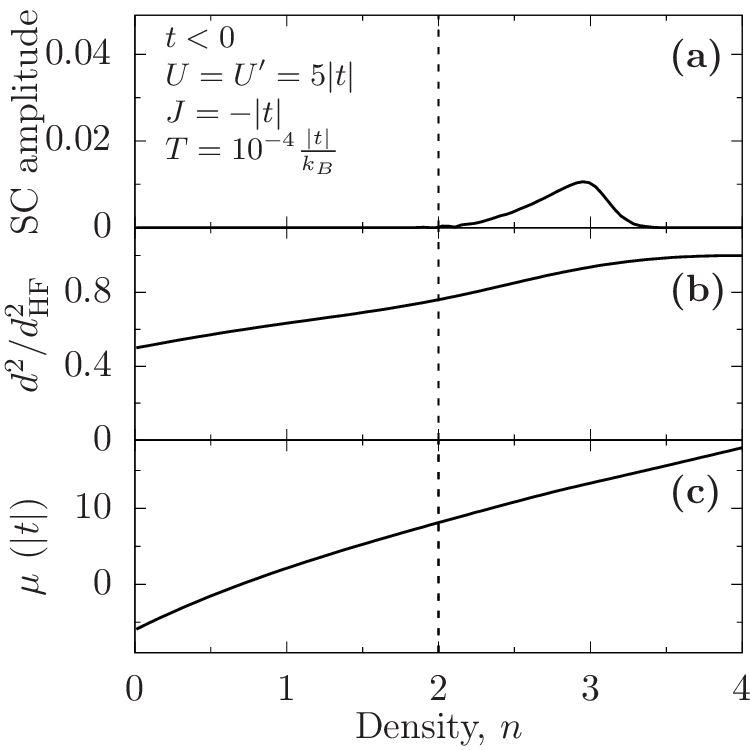} %
  \caption{Same as in Fig.~\ref{fig:phase_diagram} except for smaller value of $U = 5|t|$. The superconducting gap appears only on the overdoped side [panel (a)], and both the double occupancy (b) and the chemical potential (c) evolve in a continuous manner.}
  \label{fig:phase_diagram_u5}
\end{figure}

For completeness, we plot in Fig.~\ref{fig:phase_diagram_u5} the principal characteristics for $U = 5|t|$, i.e., in the weak-correlation limit. There is no sign of the Mott insulating behavior, so indeed appreciable correlations are required to reproduce experimental-data trend. With the increasing $U$, a second dome appears and becomes predominant at large $U$ (cf. Fig. \ref{fig:phase_diagram}).

\section{Topological properties}
\label{sec:topological_properties}

It is established that $d+id$ pairing symmetry might render the system a topological SC.\cite{ChernAIPAdvances2016} For the model~\eqref{eq:hamiltonian_final} we have explicitly computed the Chern number by the efficient method of Brillouin-zone triangulation \cite{FukuiJPhysSocJapan2005} with the result $\pm 4$, depending on the direction of phase winding of the $d+id$ order parameter. In this situation, a distinct set of topologically-protected edge states is expected for finite-size sample. To investigate the latter we have considered the lattice slab of dimensions $40\times 256$ sites with  open- and periodic-boundary conditions along the shorter and longer ends, respectively. This results in the $80$-band, one-dimensional system. The parameters were set to the same values as in previous section and the electron concentration was fixed at $n = 2.4$ to stay away from the phase-separation regime that would hinder the analysis (cf. Fig.~\ref{fig:phase_diagram}).

\begin{figure}
  \includegraphics[width = \columnwidth]{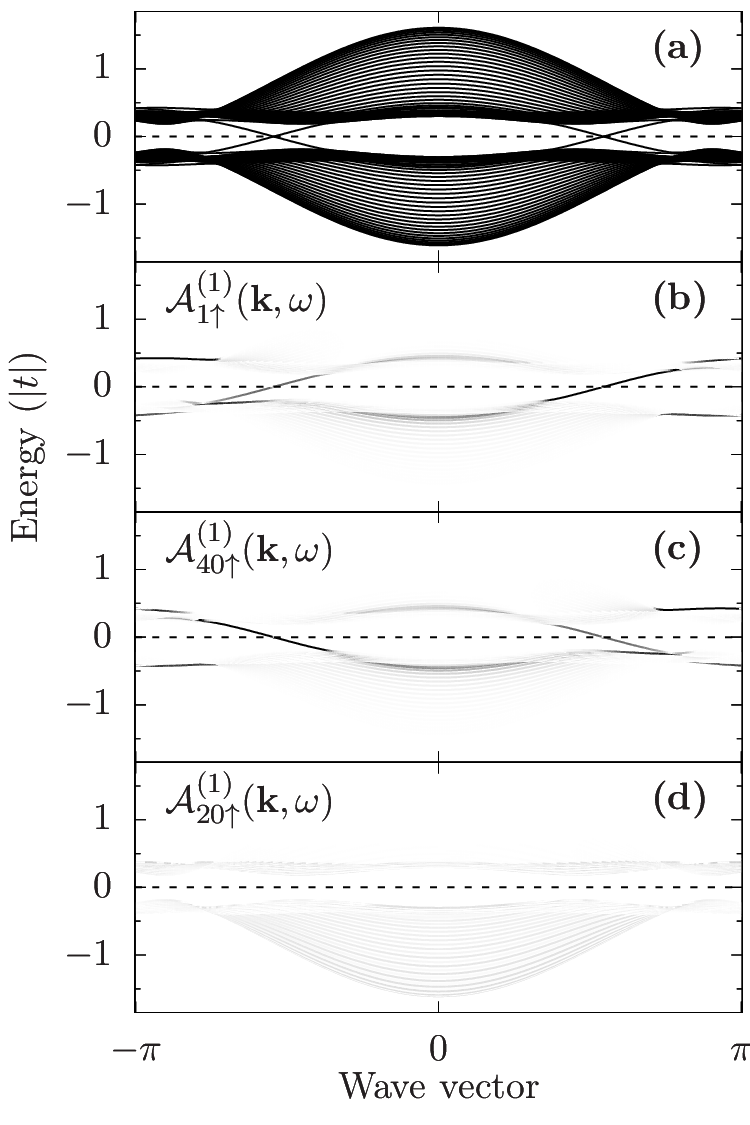} %
  \caption{Band structure and spectral properties of a $40 \times 256$ lattice slab with periodic and open boundary conditions along the longer and shorter edges, respectively. The model parameters are $U = U' = 18|t|$, $J=-|t|$, and temperature $T = 10^{-4}|t|/k_B$ ($t < 0$). (a) Band structure presented in the one-dimensional Brillouin zone. The bulk gap arises due to $d + id$ SC, whereas the levels crossing the Fermi energy originate from the topologically-protected edge states. (b)-(c) Spectral functions calculated for the orbitals on two opposite shorter ends of the sample. Note that they contribute substantially to the levels crossing the Fermi energy. (d) The same as in (b)-(c), but for the bulk states at the system center. The varying intensity reflects the difference in the spectral density values.}
  \label{fig:topo}
\end{figure}

\begin{figure}
  \includegraphics[width = \columnwidth]{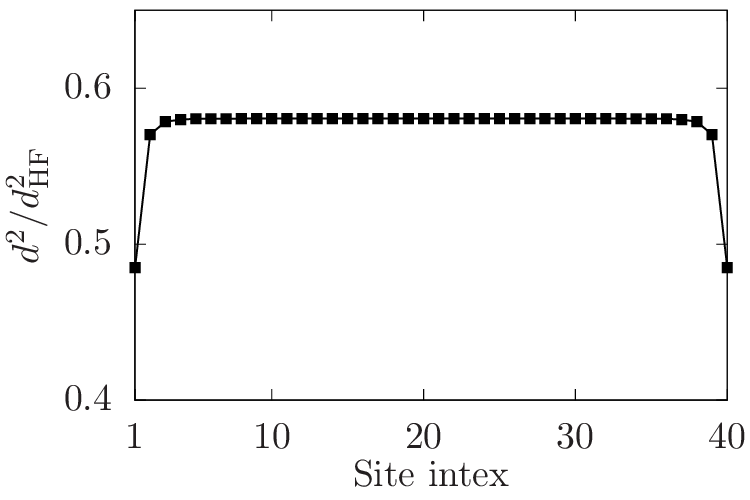} %
  \caption{Probability of the double occupancy of the lattice site $d^2$, normalized to its Hartree-Fock value $d^2_\mathrm{HF}$ across the transverse cross-section of the $40 \times 256$ lattice slab (the sites are enumerated from $1$ to $40$). The model parameters are $U = U' = 18|t|$, $J=-|t|$, temperature $T = 10^{-4}|t|/k_B$, and $t < 0$.}
  \label{fig:d2}
\end{figure}

In Fig.~\ref{fig:topo}(a) we plot the band structure of the effective quasiparticles, calculated within the SGA approach. Most of the bands are gapped due the $d+id$ superconductivity, but gapless modes crossing the Fermi level appear as well. To elucidate the nature of those states, we calculate the zero-temperature spectral functions

\begin{align}
 \frac{\mathcal{A}^{(l)}_{j\sigma}(\mathbf{k}, \omega)}{2\pi} = &  \sum \limits_n \delta(\omega - E_n + E_0) \times |\langle n|c^{(l)\dagger}_{\mathbf{k}j\sigma}|0\rangle|^2 + \nonumber \\ & \sum \limits_n \delta(\omega + E_n - E_0) \times |\langle 0|c^{(l)\dagger}_{\mathbf{k}j\sigma}|n\rangle|^2,
\end{align}

\noindent
where the index $j = 1, \ldots, 40$ enumerates the sites along the shorter edge of the system ($j = 1$ and $j = 40$ correspond to the two opposite sides of the sample). The operator $c^{(l)\dagger}_{\mathbf{k}j\sigma}$ creates a spin-$\sigma$ electron in the $l$-th orbital at the site $j$. The one-dimensional wave vector $\mathbf{k}$ results from periodic boundary conditions in the longer direction. Moreover, $E_n$ denote eigenvalues of the renormalized one-particle Hamiltonian emerging within the SGA calculations. In Fig.~\ref{fig:topo}(b) and (c) we plot the spectral function for the $l = 1$, spin-up electrons at two opposite ends of the sample, $\mathcal{A}^{(1)}_{1\uparrow}(\mathbf{k}, \omega)$ and $\mathcal{A}^{(1)}_{40\uparrow}(\mathbf{k}, \omega)$, respectively. It is apparent that the edges of the system contribute substantially to the states crossing the Fermi energy, with the opposite signs of the Fermi-velocities. The nature of these modes is finally settled in panel (d), where we display the bulk contribution to the spectral weight $\mathcal{A}^{(1)}_{20\uparrow}(\mathbf{k}, \omega)$. The latter exhibits no intensity close to Fermi energy.

Finally, we address the evolution of the electronic correlations as one moves from the edge to the sample bulk region. In Fig.~\ref{fig:d2} we plot the probability of site double occupancy $d^2$ normalized to its Hartree-Fock value as a function of the site index $j = 1, \ldots, 40$. In the central part of the system, $d^2/d^2_\mathrm{HF}$ remains practically constant. At the edges, where the topologically protected states are located, $d^2/d^2_\mathrm{HF}$ is substantially reduced. This implies that, even though the edge states are robust, it is energetically beneficial to suppress their weight by strongly correlating orbitals near the boundary of the system.

\section{Conclusions}
\label{sec:summary}

We have provided a semiquantitative analysis of the two-orbital model of bilayer graphene in the limit of extremely narrow bands. The onset of the Mott insulating state at the half-filling requires the presence of relatively strong correlations, here $U = 18|t|$. The absence of the electron-hole symmetry for the triangular lattice assumed here leads to the asymmetric SC domes on the electron and hole sides, with more prominent character on the low band-filling side. This feature differs from that in the case of two-dimensional model of high-temperature SC. \cite{ZegrodnikPhysRevB2017,Abram2017} The topological states appear naturally in the gapful $d+id$ state of the superconductor. Further studies would require explicit inclusion of electron-concentration dependence (and  sign reversal) of the effective intersite exchange integral, which should lead to spin-singlet pairing with essentially the same type of behavior. Also, several recent studies, based on the group-theoretic arguments and explicit construction of the Wannier functions,\cite{ChunArXiv2018,KangArXiv2018,KoshinoArXiv2018} suggest that the appropriate effective models for the narrow bands should be placed on honeycomb rather than triangular lattice in order to account for symmetry-related band features close to charge neutrality. Moreover, for such a lattice, it is justifiable to neglect the presence of antiferromagnetic either spin or orbital ordering. Simply put,  the model considered here is intended to account for the physics of the lower narrow bands and thus it does not rely on the details of band structure close to charge neutrality point. Keeping that in that in mind, we have included the effects of strong correlations in the flat bands, as well as second order kinetic exchange processes that turn out to be essential to the emergence of the SC state. In this sense, the obtained results concerning SC are expected to exhibit a degree of universality and be robust to model refinement (parenthetically, this is supported by apparent similarities between TBG and the high-$T_c$ cuprates). An essential extension of the present analysis would be to employ the treatment discussed here to a microscopically-derived effective model of the TBG.

\textit{Acknowledgments---}The discussions with Adam Rycerz and Micha{\l} Nowak are gratefully acknowledged. This work was supported financially by the Grant No. DEC-2012/04/A/ST3/00342  from  National Science Centre (NCN) of Poland.

\appendix

\section{Estimate of the kinetic exchange integrals}
\label{appendix:exchange_integrals}

It is  well established that two-band systems in the strong-correlation limit exhibit exhibit antiferromagnetic order at quarter filling ($n = 1$). and antiferromagnetic ordering at the half-filling, reflecting the canonical Mott insulating state.\cite{SpalekJPhysC1980,ChaoPhysStatSolidiB1977} Here we address this effect for the $\mathrm{SU}(4)$ model \eqref{eq:hamiltonian} with $U  = U'$, extended by the symmetry-breaking Hund's-type interaction $\mathcal{H}_{H} \propto - J_H \sum_i \mathbf{S}^{(1)}_i  \mathbf{S}^{(2)}_i$ with $J_H \ll U$. We show explicitly that in the doping range $n \sim 1$, ferromagnetic intersite correlations are preferred in this situation. As one approaches half-filling ($n \rightarrow 2$), intersite spin singlets become supported as a non-trivial consequence of ferromagnetic intrasite interactions. This effect might be relevant to TBG by inducing the change of pairing symmetry from triplet to singlet at the critical doping $1 \leq n_c \leq 2$.

First, note that the local charging of the site sets the dominant energy scale for the system $\sim U/2 \cdot  n_i (n_i - 1) $. Specifically, the cost due to $U$ of moving an electron from site $i$ to $j$ reads $U/2 \cdot  (n_j+1)n_j + U/2 \cdot  (n_i-1)(n_i-2) - U/2 \cdot  n_j (n_j - 1) - U/2 \cdot  n_i (n_i - 1) = U (1 + n_j - n_i)$, i.e., it can be considered a low-energy process only if $n_j = n_i - 1$. Those low-energy charge transfer gives the rise to the residual hopping. On the other hand, the high-energy direct hopping with $n_j \geq n_i$ can be eliminated by a canonical transformation that we sketch below.  Explicitly, the ``raising'' operator moving the electron from the low- to the high-energy sectors takes the form

\begin{align}
  \mathcal{H^{+}} = {\sum\limits_{i j n}}' \mathcal{H}_{ij, n}^{+} = {\sum\limits_{i j n}}' \sum_{l\sigma} c^{\dagger(l)}_{i\sigma} c^{(l)}_{j\sigma} P_{n_i \geq n} P_{n_j = n},
\label{eq:raising_operator}
\end{align}

\noindent
where $P_{n_j = n}$ and $P_{n_i \geq n}$ are projection operators onto the states with occupancies $n_j = n$ and $n_i \geq n$, respectively. The prime symbol indicates summation over nearest neighbors. The related ``lowering'' operator $\mathcal{H}^{-}_{ji, n} = \mathcal{H}^{+\dagger}_{ij, n}$ is also introduced. In the following we will retain only the processes proportional to $P_{n_i = n} P_{n_j = n}$ out of the operator \eqref{eq:raising_operator}. This is well justified as the remaining ones require substantially larger energies (by at least $U$) and thus will contribute less to the kinetic exchange. Nonetheless, in principle, it is straightforward yet tedious to include the latter as well. 

The second-order two-site kinetic exchange processes are then evaluated in a standard manner \cite{SpalekJPhysC1980,ChaoPhysStatSolidiB1977}, yielding the Hamiltonian

\begin{align}
  \mathcal{H}_\mathrm{ex} =& -\frac{1}{2}{\sum_{ijn}}' \mathcal{H}^{-}_{ji, n}\sum_{\alpha\beta\gamma\rho}\frac{P_i^{(\alpha)} P_j^{(\beta)} \mathcal{H}^{+}_{ij, n} P_i^{(\gamma)} P_j^{(\rho)}}{\epsilon^\alpha + \epsilon^\beta - \epsilon^\gamma - \epsilon^\rho} - \nonumber \\ & -  \frac{1}{2}{\sum_{ijn}}' \mathcal{H}^{+}_{ji, n}\sum_{\alpha\beta\gamma\rho}\frac{P_i^{(\alpha)} P_j^{(\beta)} \mathcal{H}^{-}_{ij, n} P_i^{(\gamma)} P_j^{(\rho)}}{\epsilon^\alpha + \epsilon^\beta - \epsilon^\gamma - \epsilon^\rho} + \nonumber\\ & + \mathrm{H.c.}, \label{eq:heff_definition}
\end{align}

\noindent
where $P^{(\alpha)}_i$ are projectors onto the $\alpha$-th local many-body configuration on site $i$ (for a two-orbital model $\alpha = 1, \ldots, 16$, giving $256$ final and initial states for a two-site interaction) and $\epsilon^\alpha$ denote local energies due to site-charging contribution and Hund's rule coupling.

We first consider the special case of $J_H = 0$ and demonstrate that the interaction, given by Eq.~\eqref{eq:exchange}, is reproduced. In this case, the for non-zero contributions to Eq.~\eqref{eq:heff_definition} always reads $\epsilon^\alpha + \epsilon^\beta - \epsilon^\gamma - \epsilon^\rho = U$ for the first line and $\epsilon^\alpha + \epsilon^\beta - \epsilon^\gamma - \epsilon^\rho = -U$ for the second (if one takes only the leading contribution form the raising operator $\propto {\sum_{i j n}}' \sum_{l\sigma} c^{\dagger(l)}_{i\sigma} c^{(l)}_{j\sigma} P_{n_i = n} P_{n_j = n}$ as assumed above). Since there is no dependence on the final and initial state indices, one can make use of the property of projection operators $\sum_\alpha P^{(\alpha)}_i = 1_i$ and write

\begin{align}
  \mathcal{H}_\mathrm{ex} =& -\frac{1}{U}{\sum_{ijn}}' \mathcal{H}^{-}_{ji, n}\mathcal{H}^{+}_{ij, n} + \frac{1}{U}{\sum_{ijn}}' \mathcal{H}^{+}_{ji, n}\mathcal{H}^{-}_{ij, n}.
\label{eq:Heff_JH_0}
\end{align}

\noindent
At the first glace, it might seem that those two terms cancel out due to opposite overall signs. This is, however, not the case as the first process is proportional to $P_{n_i=n}P_{n_j=n}$, whereas the other to $P_{n_i=n-1}P_{n_j=n+1}$. Close to integer fillings, where charge is nearly quenched by correlations, the first term dominates the second. If those projection operators are handled by mean-field-type decoupling (which is a rough yet reasonable approximation), one reproduces Eq.~(\ref{eq:exchange}) with doping-dependent effective exchange $J_\mathrm{ex} \sim t^2/U$.

An important methodological remark is in place here. Whereas the functional form of Eq.~(\ref{eq:exchange}) is robust to the model details, the precise numerical value of the effective exchange integral cannot be reliably obtained within the canonical perturbation expansion applied to the effective model alone. This situation is analogous to that for high-$T_c$ superconductivity, where the one-band Hubbard models are often used to describe essential physics. Specifically, by taking the reasonable values of the Hubbard $U$ and the hopping parameter $t$, one arrives at substantially underestimated exchange integral, and extended models of the $t$-$J$-$U$ form \cite{SpalekPhysRevB2017} need to be used to capture both the correct $U$ and magnetic exchange $J$. This is caused by sensitivity of $J$ to the details of the underlying \textit{full microscopic} Hamiltonian (such as $p$-$d$ hybridization for the case of the cuprates). 

At this point we can discuss the effects of the Hund's rule coupling $J_H$. We resort to qualitative analysis due to large number of initial and final states that cannot be handled in a straightforward manner due to lowered symmetry ($256 \times 256$ configurations in total).  First, we make a simplification by disregarding the term in the second line of Eq.~\eqref{eq:heff_definition}. This is, one again, justified close to integer filling due to charge quench. Second, since $\mathcal{H}_{H}$ annihilates states with $n \geq 3$, $J_H$ will possibly show up only in the contributions containing $\mathcal{H}^{+}_{ij, n}$ with $n=1$ and $n=2$. Below we consider two cases: the vicinity of quarter-filling ($n \approx 1$) and half-filling ($n \approx 2$).

  \begin{figure}
  \includegraphics[width = \columnwidth]{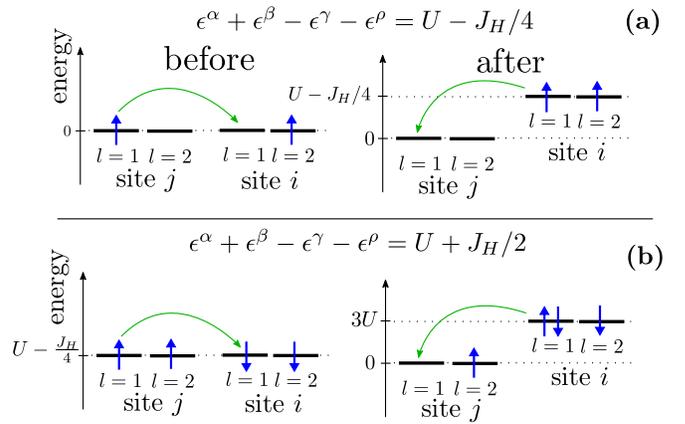} %
  \caption{Exemplary processes involving triplet states, contributing to the second-order kinetic exchange close to the quarter-filling (a), and half-filling (b). The left panels show the initial configurations of two involved sites, $i$ and $i$, whereas the right panels represent the situation after hopping takes place. In the quarter-filling case, the Hund's rule exchange reduces the energy cost of the hopping process favoring triplet configurations. On the contrary, close to the half-filling, breaking of the initial spin triplets by hopping increases the energy cost of such a process, disfavoring the triplet formation. Green arrows illustrate action of the hopping term.}
  \label{fig:processes}
\end{figure}

\subsection{Quarter-filling}

Close to quarter-filling ($n \approx 1$), the most likely initial configuration is $n_i = 1$ and $n_j = 1$. Then the Hund's rule coupling contributes differently, depending on whether the final doubly-occupied site is in singlet or triplet configuration. The denominator of the expression in the first line of Eq.~\eqref{eq:heff_definition} for $n = 1$ can be then written as $\epsilon^\alpha + \epsilon^\beta - \epsilon^\gamma - \epsilon^\rho = U + \epsilon_H^\alpha$, where $\epsilon_H^{\alpha} = -1/4 J_H$ and $3/4 J_H$ for $\alpha$ corresponding spin-triplet and singlet, respectively. The denominator takes then the smallest value for spin-triplet configuration, resulting in the largest exchange integral. This supports intersite triplet pairing. The qualitative picture behind this is simple: Hund's rule leads to splitting of the energy levels corresponding to the final local electronic configurations, reducing the energy cost of hopping to triplet configuration relative to the singlet case. This situation is illustrated schematically in Fig.~\ref{fig:processes}(a).

An estimate for the correction to the effective Hamiltonian due to $J_H$ is

\begin{align}
  \delta \mathcal{H}_\mathrm{eff} \approx \frac{1}{2} {\sum_{ i j\alpha}}' \left( \frac{1}{U} - \frac{1}{U + \epsilon_H^\alpha} \right) \mathcal{H}^{-}_{ji, 1} P^{(\alpha)}_i \mathcal{H}^{+}_{ij, 1} + \mathrm{H.c.}
  \label{eq:heff_correction_n_1}
\end{align}

\noindent
and the corresponding singlet-triplet splitting is of the order of $t^2 J_H /U^2$ with triplet having lower energy.

\subsection{Half-filling}

For $n \approx 2$, the most likely initial situation is $n_i = 2$ and $n_j = 2$. This means that, in this case, initial rather than final states are split by the Hund's coupling (that acts non-trivially only on doubly occupied sites). The denominators therefore read $\epsilon^\alpha + \epsilon^\beta - \epsilon^\gamma - \epsilon^\rho = U - \epsilon_H^\alpha - \epsilon_H^\beta$ with the definitions the same as for the quarter-filling. Note the sign change of $\epsilon_H^{\alpha, \beta}$ that implies that the largest coupling constant is now obtained for the spin-singlet configurations in the initial state. This is illustrated in Fig.~\ref{fig:processes}(b).

An estimate for the correction for $n \approx 2$ is

\begin{align}
  \delta \mathcal{H}_\mathrm{eff} \approx& \frac{1}{2} {\sum_{ i j \alpha \beta}}' \left( \frac{1}{U} - \frac{1}{U - \epsilon_H^\alpha - \epsilon_H^\beta} \right) \mathcal{H}^{-}_{ji, 2} \mathcal{H}^{+}_{ij, 2} P^{(\alpha)}_i P^{(\beta)}_j + \nonumber \\ + &\mathrm{H.c.}
  \label{eq:heff_correction_n_1}
\end{align}

The change of sign occurs at $n = n_c$, where the ferromagnetic and antiferromagnetic contributions become comparable. For the band filling $2 > n > n_c > 1$ the pairing symmetry changes thus from that of spin triplet to the singlet. It is interesting to note that there is a dip in the superconducting ordering temperature,\cite{CaoNature2018_SC} which, as we speculate, may be related to the sign change of the exchange integral. Such a change of pairing symmetry would vindicate the present real-space pairing concepts.

\section{Singlet-triplet transformation}
\label{appendix:singlet-triplet_transformation}

Here we demonstrate that by making use of the symmetry of the original model \eqref{eq:hamiltonian}, the triplet order parameter, defined by Eqs.~\eqref{eq:sc_structure_singlet}-\eqref{eq:sc_structure_d}, can be transformed into a singlet. This equivalence means, first of all, that the dependence for the singlet and triplet cases of the superconducting gap versus band filling will be qualitatively the same. Such a transformation may be relevant close to half-filling ($n=2$), where (in the presence of Hund's coupling term) the intraorbital kinetic antiferromagnetic exchange dominates interorbital one. Thus, this ability to map one situation onto another suggests general character of the obtained SC phase diagram. Nonetheless, quantitative analysis of the triplet to singlet transition with the band filling would require precise knowledge of the kinetic exchange integral scaling with the electron concentration.

We introduce a unitary matrix

\begin{align}
  \label{eq:transformation_matrix}
   \mathcal{U} =
  \left( {\begin{array}{cccc}
   1 & 0 & 0 & 0 \\
   0 & 0 & 1 & 0 \\
   0 & 1 & 0 & 0 \\
   0 & 0 & 0 & 1 \\
  \end{array} } \right)
\end{align}

\noindent
that defines new set of annihilation operators $\tilde{c} \equiv \mathcal{U} c$ with $c^\dagger \equiv (c^{(1)\dagger}_\uparrow, c^{(1)\dagger}_\downarrow, c^{(2)\dagger}_\uparrow, c^{(2)\dagger}_\downarrow)$. The latter operation is essentially a spin flip assisted by the orbital exchange. Now, the pairing operator for the the $A$-type triplet SC can be written as $\hat{\Delta}_{ij}^{\mathrm{triplet}} \equiv c^{(1)\dagger}_{i\uparrow} c^{(2)\dagger}_{j\uparrow} - c^{(2)\dagger}_{i\uparrow} c^{(1)\dagger}_{j\uparrow} + c^{(1)\dagger}_{i\downarrow} c^{(2)\dagger}_{j\downarrow} - c^{(2)\dagger}_{i\downarrow} c^{(1)\dagger}_{j\downarrow}$. By making use of the transformation \eqref{eq:transformation_matrix}, one can show explicitly that $\hat{\Delta}_{ij}^{\mathrm{triplet}} \rightarrow \hat{\Delta}_{ij}^{\mathrm{singlet}} \equiv c^{(1)\dagger}_{i\uparrow} c^{(1)\dagger}_{j\downarrow} - c^{(1)\dagger}_{i\downarrow} c^{(1)\dagger}_{j\uparrow} + c^{(2)\dagger}_{i\uparrow} c^{(2)\dagger}_{j\downarrow} - c^{(2)\dagger}_{i\downarrow} c^{(1)\dagger}_{j\uparrow}$. The triplet pairing operator thus turns into the sum of two decoupled singlet operators in each orbital channel. In the direct vicinity of the half-filling, antiferromagnetic intraorbital antiferromagnetic exchange is expected to dominate its interorbital counterpart (cf. Appendix~\ref{appendix:exchange_integrals}), providing the attractive coupling interaction in the singlet channel $\propto - \hat{\Delta}_{ij}^{\mathrm{singlet}\dagger} \cdot \hat{\Delta}_{ij}^\mathrm{singlet}$. Since the transformation of this pairing potential into the $\propto - \hat{\Delta}_{ij}^{\mathrm{triplet}\dagger} \cdot \hat{\Delta}_{ij}^\mathrm{triplet}$, considered in this paper, is simply a matter of unitary transformation $\mathcal{U}$ that does not modify the zeroth-order Hamiltonian \eqref{eq:hamiltonian} (but turns triplet pairing into singlet one), the solutions described here are relevant to both scenarios.

\section{Diagrammatic extensions}
\label{appendix:diagrams}

The SGA approximation, employed in this work, represents simple, but essential amendment to the original Gutzwiller approximation (GA), known under the acronym \textit{the renormalized mean field theory} in the context of physics of strongly-correlated systems and, in particular, of high-temperature SC (see, e.g., \citenum{EdeggerAdvPhys2007}), was required. The revision is necessary, since, within GA, the self-consistent (Bogoliubov-de Gennes) equations for the order parameter does do not coincide with the variational optimization of the system free energy with respect to it.\cite{JedrakArXiV2010} In the sense, the GA approximation violates the \textit{Bogoliubov theorem} of statistical consistency. To improve the situation, additional constraints must be imposed to fulfill the theorem.\cite{JedrakPhysRevB2011} The subsequently developed diagrammatic expansion for the Gutzwiller wave function (DE-GWF)\cite{ZegrodnikNewJPhys,Fidrysiak_kspace} Here we briefly describe this diagrammatic extension, which allows us to go beyond the SGA limit and obtain the full Gutzwiller wave function solution to a desired accuracy (without imposing the formal limit of infinite spatial dimensionality).

First, to improve the efficiency of the DE-GWF calculation scheme, the constraint

\begin{equation}
 \hat{P}_i^2\equiv1+x\hat{d}^{\textrm{HF}}_i,
 \label{eq:constraint}
\end{equation}

\noindent
is applied with respect to the correlation operator $\hat{P}_G$.\cite{Bunemann2012}  Here $x$ is a variational parameter and $\hat{d}^{\textrm{HF}}_i\equiv\hat{n}_{i\uparrow}^{\textrm{HF}}\hat{n}_{i\downarrow}^{\textrm{HF}}$, $\hat{n}_{i\sigma}^{\textrm{HF}}\equiv\hat{n}_{i\sigma}-n_{0}$, with $n_{0}\equiv\langle\Psi_0|\hat{n}_{i\sigma}|\Psi_0\rangle$. Since all the $\lambda_{\gamma}$ coefficients from $\hat{P}_G$ can be expressed by the use of $x$, we are left with only one variational parameter. The expectation value of the system energy in the correlated state $|\Psi_G\rangle$ can be expressed as
   \begin{equation}
  \langle\Psi_G|\hat{o}_{i}\hat{o}^{\prime}_{j}|\Psi_G\rangle=\sum_{k=0}^{\infty}\frac{x^k}{k!}\sideset{}{'}\sum_{l_1...l_k}\langle\Psi_0| \tilde{o}_{i}\tilde{o}^{\prime}_{j}\;\hat{d}^{\textrm{HF}}_{l_1...l_k}|\Psi_0 \rangle,
\label{eq:expansion}
\end{equation}
where $\tilde{o}_{i}\equiv\hat{P}_i\hat{o}_{i}\hat{P}_{i}$, $\tilde{o}^{\prime}_{j}\equiv\hat{P}_j\hat{o}^{\prime}_{j}\hat{P}_{j}$, $\hat{d}^{\textrm{HF}}_{l_1...l_k}\equiv\hat{d}^{\textrm{HF}}_{l_1}...\hat{d}^{\textrm{HF}}_{l_k}$, $\hat{d}^{\textrm{HF}}_{\varnothing}\equiv 1$, and $\hat{o}_i$, $\hat{o}_j$ are any two local operators from our Hamiltonian (\ref{eq:hamiltonian}). The primed summation has the restrictions $l_p\neq l_{p'}$, $l_p\neq i,j$ for all $p$ and $p'$.

Note that the expectation values on the right-hand side of Eq.~(\ref{eq:expansion}) are calculated in the non-correlated state $|\Psi_0\rangle$. This allows to apply the Wick's theorem and express the energy of the system in the correlated state in terms of the non-correlated expectation values $P_{ij}=\langle\Psi_0|{c}_{i\sigma}^{\dagger}{c}_{j\sigma}|\Psi_0\rangle$ (here we limit to the paramagnetic state -- no anomalous SC averages are included) and the variational parameter $x$. The expressions which result from the Wick's decomposition can be interpreted as diagrams for which the atomic sites have the interpretation of vertices, while the averages $P_{ij}$ play the role of lines connecting those vertices. When carrying out the calculations, the diagrams have to be summed over the lattice in real space by attaching their inner vertices [indexed by $l_1$\ldots$l_k$ in Eq.~(\ref{eq:expansion})] to the lattice-sites in all possible configurations determining the corresponding contributions to the system energy. This procedure corresponds to the summation over $l_1$\ldots$l_k$ in Eq.~(\ref{eq:expansion}). An alternative approach of $\mathbf{k}$-space summation has also been introduced recently.\cite{Fidrysiak_kspace} In practice, a real-space cut-off, $R_{\mathrm{max}}$, is introduced limiting the range within which the diagrams are summed on the lattice, as well as determining the number of different lines that have to be included in the calculations. Here, we take $R_{\mathrm{max}}=3a$, which requires including the lines $\langle\Psi_0|{c}_{i\sigma}^{\dagger}{c}_{j\sigma}|\Psi_0\rangle$ up to the fifth nearest-neighbor. Also, it is not possible from obvious reasons to carry out the summation for $k\rightarrow \infty$. However, in most cases the first 4-5 terms of the expansion in $x$ allow to reach the convergence,\cite{Abram2017} whereas the zeroth order expansion is equivalent to the SGA approach. In Fig. 7 we show an exemplary second-order diagram, which results from calculating the hopping term expectation value $\langle\Psi_G|{c}^{\dagger}_{i\sigma}{c}_{j\sigma}|\Psi_G\rangle$. The expression for the ground state energy in the state $|\Psi_G\rangle$ obtained in the described manner can further be applied to the statistically consistent calculation scheme analogous to that described in Sec.~\ref{sec:solution}.

\begin{figure}
  \includegraphics[width = 0.75\columnwidth]{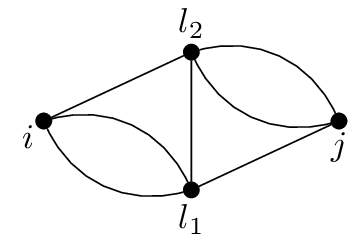} %
  \caption{Exemplary diagram resulting from the Wicks decomposition of $\langle\Psi_0|{c}^{\dagger}_{i\sigma}\hat{n}^\mathrm{HF}_{i\bar{\sigma}}{c}_{j\sigma}\hat{n}^\mathrm{HF}_{j\bar{\sigma}}\hat{d}^{\textrm{HF}}_{l_1}\hat{d}^{\textrm{HF}}_{l_2}|\Psi_0\rangle$, which is one of the four second order expansion terms from (\ref{eq:expansion}), when $\hat{o}_i={c}^{\dagger}_{i\sigma}$ and $\hat{o}'_j={c}_{j\sigma}$ (i.e., when calculating the hopping contribution in the correlated state $\langle\Psi_G|{c}_{i\sigma}^{\dagger}{c}_{j\sigma}|\Psi_G\rangle$). The black lines connecting the vertices correspond to $P_{mn}$ expectation values. During the summation procedure in real space the so-called internal vertices $l_1$ and $l_2$ are attached to all possible lattice sites within the region determined by the real-space cut-off $R_{\mathrm{max}}$ ($|\mathbf{R}_{l}-\mathbf{R}_i|\leq R_{\mathrm{max}}$ and $|\mathbf{R}_{l}-\mathbf{R}_j|\leq R_{\mathrm{max}}$  ). }

  \label{fig:diagram_T33}
\end{figure}%

\begin{figure}
  \includegraphics[width = \columnwidth]{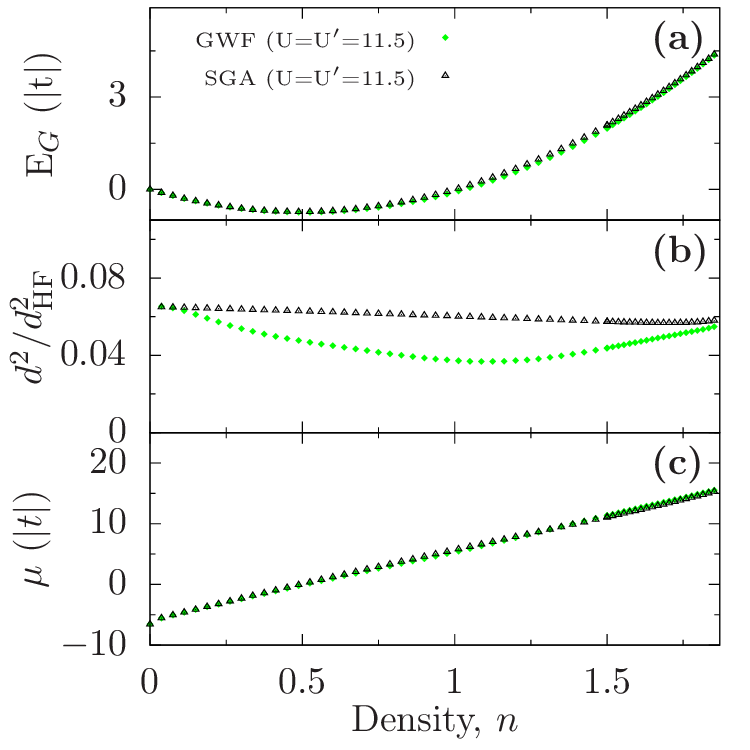} %
  \caption{Ground state energy (a), double occupancies (b), and chemical potential (c), all as a function of band filling for the case of two-band Hubbard model with $U=U'=11.5|t|$ (cf. Eq.~\ref{eq:hamiltonian}) and $J = 0$, calculated within the SGA (red squares and black triangles) and DE-GWF methods (blue dots and green diamonds, respectively). Only the paramagnetic phase has been included (no SC phase considered here). The DE-GWF calculation has been carried out up to the third diagrammatic order.}
  \label{fig:de-gwf_results}
\end{figure}%

\begin{figure}
  \includegraphics[width = \columnwidth]{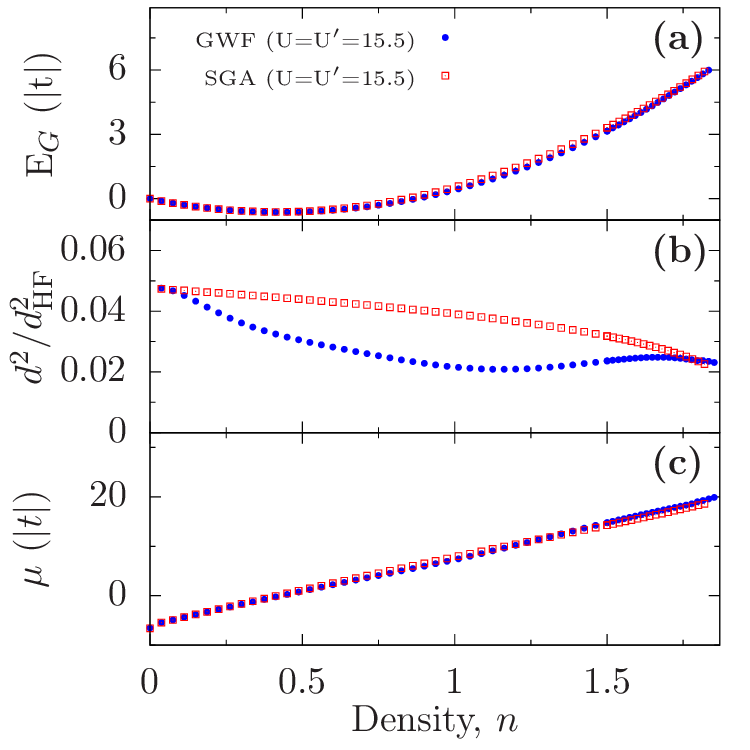} %
  \caption{The same as in Fig.~\ref{fig:de-gwf_results}, but for $U=U'=15.5|t|$.}
  \label{fig:de-gwf_results_2}
\end{figure}%

In Figs.~\ref{fig:de-gwf_results} and \ref{fig:de-gwf_results_2} we show the comparison between the SGA results with those corresponding to the third order DE-GWF calculations for the case of the two-band Hubbard model given by Eq.~(\ref{eq:hamiltonian}) with two selected values of $U=U'$. As one can see the global quantities such as ground state energy and chemical potential [panels (a) and (c)] are almost identical for both methods. Differences can be seen for the case of double occupancies especially close to $n\approx 1$. We show the results for the band fillings up to $n\approx 1.85$, since for the region close to the half-filled situation problems with the convergence appeared when carrying out the DE-GWF calculations. Nevertheless, it can be seen that in the most interesting for us regime which is above $n\approx 1.5$ the double occupancies calculated within SGA and DE-GWF converge, which justifies the choice of the simpler approach (SGA) in the extended analysis carried out in Secs.~\ref{sec:model}-~\ref{sec:summary}. Detailed discussion of the higher-order effects will be provided separately.

\bibliography{bibliography}

\end{document}